\documentclass[iop]{emulateapj}

\begin{document}

\title{Modern Cosmology: Assumptions and Limits}
\author{Jai-chan Hwang${}^{1,2}$}
\address{${}^{1}$Department of Astronomy and Atmospheric Sciences,
                Kyungpook National University, Daegu 702-701, Republic of Korea \\
         ${}^{2}$Korea Institute for Advanced Study, Seoul 130-722, Republic of Korea}


\begin{abstract}
Physical cosmology tries to understand the Universe at large with its origin and evolution. Observational and experimental situations in cosmology do not allow us to proceed purely based on the empirical means. We examine in which sense our cosmological assumptions in fact have shaped our current cosmological worldview with consequent inevitable limits. Cosmology, as other branches of science and knowledge, is a construct of human imagination reflecting the popular belief system of the era. The question at issue deserves further philosophic discussions. In Whitehead's words, ``philosophy, in one of its functions, is the critic of cosmologies.'' (Whitehead 1925)
\end{abstract}


%
%
%
\section{Science and Cosmology}

Alexander Calder has mentioned that ``the universe is real but you can't see it, you have to imagine it.'' 
As an artist Calder's universe may mean everything in the world. Modern physical cosmology aims at understanding quantitatively the structure, origin and evolution (sometimes including future) of the Universe based on scientific methods (i.e., observation and experiment). To scientists, the Universe is a physical object\footnote{Professor Roberto Torretti objects calling the Universe as a ``physical object''. According to him, ``the epistemic counterpart to artist Calder's dictum would be something like: `The universe is real, but you can't grasp it as an object, you have to think of it as a Kantian Idea'.'' (Torretti 2011)} in the large-scale. However, regions in the Universe accessible by observation and experiment are limited. The forbidden regions include outside our current cosmic horizon, time already passed in our past light-cone, and the far future which has yet to come. Some of these regions are not just practically difficult to access, but they belong to the absolute limit of scientific knowledge which is in principle inaccessible by observation or experiment.

Of course, science is not simply based on observation and experiment so innocently. If science were naively based on observation and experiment, the science as we know now may not be possible. On the contrary, science in reality is more related with the art of ignoring and selecting observations, and manipulating experiments, in accordance with a preconceived theory. Detailed observation is often a hindrance to scientific reasoning. Ignore apparent phenomena and grasp the essence. Thus, in science theory often comes before observation. The trick is to treat the subject as an isolated, simplified, idealized and abstract (preferably mathematized) model, and to test and materialize it by fitting data to a model using the method of analysis and statistical techniques. In this way, the individuality is lost. Modern cosmology shows such a state of affairs well. After all, scientific cosmology is nothing other than `the Universe imagined based on the ``scientific'' method.' In fact, the situation of cosmology is not alone in science. In Albert Einstein's words, ``physical concepts are free creations of the human mind, and are not, however it may seem, uniquely determined by the external world.''

Although science is an effort to approximate phenomena through models and to test those models, it is important to be aware of the difference between model and reality. As remarked by Alfred North Whitehead, ``the aim of science is to seek the simplest explanations of complex facts. We are apt to fall into the error of thinking that the facts are simple because simplicity is the goal of our quest.'' (Whitehead 1920)

%
%
%
\section{Cosmological Principle}

In cosmology, in order to infer regions unknowable even in principle, an assumption is necessary, which might reflect our anticipation. An assumption that distribution of matter in the large-scale is spatially homogeneous and isotropic is termed the cosmological principle, often adopted in modern physical cosmology. Its origin can be traced to Einstein's paper in 1917, in which he has applied his newly introduced gravity theory to cosmology in order to reconcile his theory with Mach's principle (Torretti 2000). Although even the presence of external galaxies was not known at that time, Einstein assumed homogeneity and isotropy of space, and thus of the matter distribution, merely for the sake of mathematical simplicity. Notice that this is an assumption not based on what was observed. Perhaps Einstein did not expect that this simple working assumption would have become a basic principle in cosmology in future development.

In fact, the basic tenet of physics is that the laws of physics we know are valid always and everywhere. Thus, advocating the universality of the laws of physics reminds us sort of the cosmological principle; this belief in universality is in general not testable. It is amusing to notice that assuming the general validity of the physical laws in space and time is more similar to the perfect cosmological principle, which adds the time invariance of the physical state of the Universe in addition to the cosmological principle. However, the perfect cosmological principle implemented in the steady-state theory is no longer popular in modern cosmology.

Although in history the cosmological principle was initially postulated theoretically without any reference to observations, one might wonder whether the assumption on matter distribution can be tested through the observations. However, there are difficulties in practice, and often in principle. There can be some evidence of isotropy around us, but the test of homogeneity becomes difficult as the scale increases. In addition, if we consider the finite speed of light, even in a perfect observation, we cannot prove the homogeneity of space. Only through models we can agree on its plausibility. The observed two-dimensional projected isotropy of the cosmic microwave background radiation does not necessarily imply that the three-dimensional matter distribution is also isotropic. Furthermore, examination beyond the horizon (light propagation distance during the age of the Universe) is in principle impossible, and the cosmological principle in those regions remains as an untestable assumption. As emphasized by George F. R. Ellis, ``the problem [is that] there is only one universe to be observed, and we effectively can only observe it from one space-time point. ¡¦ Given this situation, we are unable to obtain a model of the Universe without some specifically cosmological assumptions which are completely unverifiable.'' (Ellis 1975)
In modern cosmology, the assumption refers to the cosmological principle.

Theoretically, without the cosmological principle, physical cosmology becomes mathematically too complicated to handle, and this practical difficulty might have an important role in accepting this simple assumption. Martin Rees has mentioned that ``principles in cosmology have often connoted assumptions unsupported by evidence, but without which the subject can make no progress.''

The cosmological principle still in large measure is a philosophical assumption, and is not based on observations or experiments as is often emphasized by the scientific method. Making assumptions is not a problem. It is fine, as long as we are aware of this nature, and try to examine the case in regions where testing is possible. From the observational side, efforts on tests of whether the Universe is homogeneous (homogeneity measure) and isotropic (isotropy measure), and whether deviations from the homogeneity-isotropy are acceptable (linearity measure), should be continued. Only in this way, physical cosmology could be defined as a {\it genuine} science.

The real problem is that the cosmological principle, which is merely our own assumption, gives a strong constraint on our perspective towards the Universe; as a consequence, it has a significant impact on the interpretation of observed results, and even on the observational strategy. In fact, this has determined our present cosmological worldview. Remarkable examples are episodes concerning the expansion model, the dark matter, and the dark energy interpretations. Let's examine these three cases.

%
%
%
\section{Impacts}

The only motion allowed under the assumption of spatial homogeneity-isotropy is the Hubble-like (velocity proportional to distance) expansion or contraction. We don't even need any theory for that. The assumption demands the result. However, current observations only tell us that the redshift increases roughly proportionally to the distance, and do not tell that the redshift is due to Doppler or cosmological expansion. In particular, in order to measure the expansion rate, an individual observation is merely aligned along the radial axis, ignoring any potential dependence of the rate on the angular direction of the object in the sky. This is surely due to the influence of the cosmological principle. Thus, even the observation is performed under direct influence of the theory. In this case, the theory is nothing more than our assumed cosmological principle.

Motions of galaxies in clusters and the rotation speed of disks in spiral galaxies are known to be too fast to be bounded by luminous matter. Without substantial amount of non-luminous matter present, galaxies and clusters are unstable and could be transient phenomena. Such non-luminous matter, only known through gravity, is termed as `dark matter'. Such an interpretation, however, is based on the assumption that Newton's (Einstein's as well) gravity is valid in galactic and cluster scales. This reflects our belief in the universality of physical laws, and particularly our faith in Newton's theory, which lead us to such a conclusion. But Newton's theory has never been tested on those scales; nor Einstein's gravity has yet been tested in cosmology. Therefore, dark matter is also a case where our belief system has affected the interpretation of the observed results.

Einstein's gravity is widely accepted as the gravity to handle astronomical phenomena. The theory holds a remarkable track record in the solar-system test based on vacuum Schwarzschild solution and the parameterized post-Newtonian approximation where the gravitational fields are supposed to be weak. Although it is true that Einstein's theory has not failed in any experimental test based on modern scientific and technological development up till today, it is also true that there has been no experimental test of the theory in the strong gravitational field and in large scale even including the galactic scale. Cosmological application of Einstein's theory requires $10^{15}$ factor (horizon scale divided by an astronomical unit) extrapolation compared with the experimentally tested scale, which is surely a staggering extrapolation. Einstein's gravity is generally accepted in cosmology mainly based on its successes in other astronomical and Earth bound tests and the theory's own prestige associated with Einstein's fame and historical legacy. Thus, Einstein's gravity can be regarded as another important assumption often adopted in cosmology.

Near the end of the last millennium, using the corrected luminosity of Type-Ia supernovae as a distance indicator, reports were made that the expansion is accelerating in time. In order to have accelerated expansion, gravity in the large-scale must have a repulsive nature. Historically, Einstein (1917) has introduced the cosmological constant $\Lambda$, which shows a repulsive nature for a positive sign, in order to achieve a consistent static model with spherical geometry. The possibility of discovering a repulsive nature of gravity in the cosmic scale is still a surprising claim. Acceleration does not necessarily imply a cosmological constant, and the agent which causes the recent acceleration is more generally termed as `dark energy'.

However, the interpretation of the observations as a presence of acceleration is based on the assumption that the matter distribution up to the observed supernovae distances and beyond is well approximated by the cosmological principle. That is, as we strive to fit the observed data (nature) within a preconceived theoretical model (theory) based on the cosmological principle, the observation is interpreted as acceleration. The situation is consistent with Kuhn's (1962) interpretation of the behavior of normal science within a paradigm.

The nature of dark matter and the nature of dark energy are regarded as two important mysteries in modern physics. Compared with the case in galaxy and cluster motions where a new matter distribution is preferred to changing the gravity theory, it is ironic to notice that in the wake of supernovae observations researchers prefer to take radical positions, rather granting the acceleration by changing gravity theories than reconsidering our basic assumption on the matter distribution. Research feasibility has certainly played a role in such diverging trends. Abandoning the homogeneity-isotropy assumption and confronting with nonlinear phenomena is a limit and a challenge faced, not only in cosmology but also in the whole science.

%
%
%
\section{Standard Model}

Despite weaknesses from the observational side, the cosmological model based on the cosmological principle is still in good shape. A reason is that the model is widely regarded as capable of providing an overall theoretical paradigm, which can consistently explain various observations. Recently, the cosmological model based on Einstein's gravity and the cosmological principle is making a show of constraining cosmological parameters with a few percent precision level through the observations, thus hailing a precision cosmology era (precision of unknowns?). The model is termed as the standard (concordance) cosmological model. The $\Lambda$CDM (cosmological constant plus cold dark matter) with `inflation' (early acceleration phase) is another name of the model. Here, dark matter, dark energy, and inflation, together with Einstein's gravity and the cosmological principle, are taken for granted; notice that the latter two are not often even mentioned.

However, whether the theory is successful or not belongs to the eye of the beholder. There is an irony in the claim that the standard cosmological model is successful in explaining all the cosmological data. All the three terms (inflation, $\Lambda$, and CDM) describing the standard model are nothing more than names referring to unknown theoretical mysteries introduced as saviors of the favored model. Here we point out that in the standard model only about 0.5\% of required agents in energy are available in light (Persic \& Salucci 1992); the other 99.5\% are unseen theoretical devices only introduced to fill the gap (calling missing 99.5\% as a gap is not an entirely fair expression). Such a tremendous effort undertaken only to maintain existing theory is likely to be unprecedented in other scientific fields.

Inflation, which is another flexible theoretical tool, was introduced to reduce the internal inconsistency of the Big-Bang model and to generate the initial seeds for later development of large-scale structures. However, if we consider its energy scale, its experimental verification is simply not feasible; in the laboratory, we are now barely reaching one tera electon-volt energy scale, and to reach inflation we need yet another factor of tera ($10^{12}$) in energy scale, which is an utter impossibility.

If the standard model means orthodox, it may also reflect well the current situation of modern cosmology. This shows an unhealthy situation, where through elimination of competing theories the standard model becomes a dogma and now presses other remaining and arising alternatives as heresy. Considering that cosmology has only a handful of observational facts to face with, and remembering that it is based on presuppositions (many unverifiable) without proper observational evidence, various alternative theories in regions where the observation has not yet reached will have neutralizing roles in our current single vision on the Universe.

What would happen if we change our assumption on the cosmological principle? This fundamentally important problem has not been pursued much, due to the mathematical difficulty in handling nonlinear processes. Not having alternatives due to practical difficulties in research does not guarantee that the presently available explanation should be correct. Scientific cosmology has the Universe as an external reality to compare. If the Universe looks simple in the standard model, is it due to its intrinsic nature or to our simplifying assumptions?

Meanwhile, if the cosmological principle is even approximately true, there must be a reason for the fact. Inflation provides a possibility that the region within our present horizon has originated from inside the horizon before the early acceleration phase, thus opening a possibility that the cosmological principle can be achieved through a causal mechanism. However, inflation has not provided the actual mechanism for that achievement. Although a class of homogeneous but anisotropic models is known to be driven to an isotropic model through acceleration, 
the general mechanism is not known yet. Why is the Universe spatially homogeneous and isotropic? This may deserve to be one of the important theoretical problems in scientific cosmology.

Dark matter, dark energy, and inflation are essential theoretical devices introduced in standard cosmology. By future observations and experiments, these may turn out to be the success of simple theoretical inferences. Or, as we encounter better theoretical alternatives, these may as well turn out to be mere theoretical devices like epicycles in Ptolemaeus' Earth-centered cosmology. That is, these three devices could be merely {\it ad hoc} concepts introduced in standard cosmology, in order to accommodate new observational results in the currently popular cosmological paradigm. In fact, the introduction of unseen dark matter has precedent cases in the history of astronomy. Here we have both success and failure stories. The discovery of Neptune as dark matter to explain the anomalous orbit of Uranus, and the attempt to explain the unaccounted precession of Mercury's perihelion by unseen dark matter Vulcan are two episodes corresponding to the success and failure, respectively, of Newton's gravity.

%
%
%
\section{Metaphysical Questions}

Concerning the Universe's future evolution, who can tell the cosmic future? If we take a specific cosmological model, the future of that model can be determined. However, how do we know that a specific model is suitable for the purpose? Besides, situations in the early stage of the Big-Bang where the energy scale is beyond our experimental reach, and regions beyond our present horizon, can be regarded as metaphysical (beyond the scope of physical science in Aristotle's sense, see Trusted 1991) domains not only practically but also in principle impossible to reach, thus demarcating the boundary of absolute scientific knowledge possible. The cosmic future, together with the early Universe and beyond the horizon, is yet another {\it Terra Incognita}. With no potential observational and experimental tests, all theoretical attempts are only explanatory arguments and cannot be distinguished from metaphysical speculations or myth. ``Un-testable science'' is an oxymoron.

Let us examine the early Universe, where no experiment will ever be possible in the foreseeable future. Did the expansion have a beginning? If there was a beginning, what does that mean? Is it possible that we have a collapsing phase before the expansion? At present, all these questions are concerned with the unknown territory, regions unreachable by observations or experiments, in other words, a province of metaphysics. However, these are questions concerning real phenomena. Therefore, one cannot deny that, as the observational and experimental range of science is expanded, there may be a chance that answers will eventually come by science in the future. Thus, these questions differ from fundamentally metaphysical questions in cosmology.

The fundamentally metaphysical cosmological questions are the following. Is the existence of the Universe necessary? What is the ultimate reason for its existence? Does the Universe have a purpose? What is the meaning of the Universe? What is the man's status in the Universe? ¡¦ The following questions are commonly raised in both physical and metaphysical cosmologies. What is the origin of the Universe? What is the ultimate building block of the Universe? Although it may look like that these questions are within scientific reach, answers to these questions are still under groping in the dark in astronomy and physics. Perhaps, as always has been the case in the history of knowledge, we may have answers which only reflect the ideology of the modern era (the science); these constitute a modern myth.

%
%
%
\section{Limits}

Gautama Buddha has remained silent on two cosmological questions. These are questions about `temporal and spatial finiteness or infiniteness' of the Universe. (These correspond to eight in the `fourteen unanswerable questions'. They are eight because, for example, for the time we have the following four possibilities: `is the world eternal?', `or not?', `or both?', and `or neither'.) Buddha has undeclared on these questions, as being metaphysical speculations irrelevant to attain the liberation and to reach nirvana, and he has discouraged his disciples from wasting time and energy on those points (Pali Canon). As these questions are related to actual facts with potential answers, we might anticipate that the answers can be reached through science. However, even with stunning scientific and technological endeavor and advances, even after 2,500 years have passed, the answers to these questions in modern physical cosmology are still unknown.

From a standpoint of standard cosmology, the answers to these questions are not merely practically difficult to unravel, but rather `impossible to answer in principle'. That is, the answers to these questions are beyond the scope of observations and experiments, thus they belong to the territory of metaphysical speculations. Of course, this is the perspective from the modern physical cosmology, and there is no reason to believe that we have finally reached the ultimate stage in our cosmological knowledge.

Such a fundamental limitation in our understanding of the Universe is not necessarily a disappointment. We may get a comfort from the following insight according to Aristotle, ``the charm and importance of a study of the Heavens was matched only by the uncertainty of the knowledge produced.'' (Aristotle)

Here are some more related wisdoms. Bertrand Russell: ``Science is what you know, philosophy is what you don't know.'' William James: ``Our science is a drop, our ignorance a sea.'' Samuel Butler: ``Science, after all, is only an expression for our ignorance of our own ignorance.'' Alfred North Whitehead: ``Not ignorance, but ignorance of ignorance, is the death of knowledge.'' Benjamin Disraeli: ``To be conscious that you are ignorant is a great step to knowledge.'' Confucius: ``To know that we know what we know, and that we do not know what we do not know, that is true knowledge.'' Then, here is the well-known wisdom by Socrates: ``I know only that I do not know,'' and that ``true wisdom lies in knowing the limits of wisdom.'' Similar statement was made by Immanuel Kant: ``It is precisely in knowing its limits that philosophy consists.'' Scientific cosmology needs the help of philosophical introspection.

%
%
\vskip .2cm
We wish to thank Professor Roberto Torretti for insightful correspondences and clarifying remarks on many issues. We also wish to thank Dr.\ Graziano Rossi for his invitation to write, careful examination of the manuscript as well as disagreements. Most of Dr.\ Rossi's disagreements remain which is natural in our view of the subject. As M.\ Rosemary Wright writes, ``cosmology itself, like all arts and sciences, is a construct of human intelligence, subject to social and linguistic conditioning and dubious means of communication.'' (Wright 1995)

%
%

\end{document}